\documentclass[conference]{IEEEtran}
\usepackage{cite}
\usepackage{amsmath,amssymb,amsfonts}
\usepackage{algorithmic}
\usepackage{graphicx}
\usepackage{textcomp}
\usepackage{xcolor}
\usepackage[hyphens]{url}
\usepackage{multirow}
\usepackage{adjustbox}
\usepackage{xspace}
\usepackage{booktabs}
\usepackage{array}
\usepackage{xfrac}
\usepackage{hhline}
\usepackage{balance}
\usepackage[all]{nowidow}
\usepackage[bookmarks=true,breaklinks=true,letterpaper=true,colorlinks,citecolor=blue,linkcolor=blue,urlcolor=blue]{hyperref}

\def\BibTeX{{\rm B\kern-.05em{\sc i\kern-.025em b}\kern-.08em
    T\kern-.1667em\lower.7ex\hbox{E}\kern-.125emX}}

\newcommand{\tbt}{{TBT}\xspace}
\newcommand{\ttft}{{T2FT}\xspace}
\newcommand{\ete}{{E2E}\xspace}

\newcommand{\deconlystage}{{decoding-only stage}\xspace}
\newcommand{\mixedstage}{{mixed stage}\xspace}

\newcommand{\duplex}{{Duplex}\xspace}
\newcommand{\bankbundle}{{bank bundle}\xspace}
\newcommand{\highp}{{xPU}\xspace}
\newcommand{\lowp}{{Logic-PIM}\xspace}
\newcommand{\bankpim}{{Bank-PIM}\xspace}
\newcommand{\bgpim}{{Bankgroup-PIM}\xspace}

\newcommand{\opb}{{Op/B}\xspace}

\newcommand{\gpu}{{GPU}\xspace}
\newcommand{\doublegpu}{{2$\times$GPU}\xspace}

\newcommand{\duplexpe}{{Duplex+PE}\xspace}
\newcommand{\duplexpet}{{Duplex+PE+ET}\xspace}

\newcommand{\duplexbankpim}{{Bank-PIM}\xspace}
\newcommand{\duplexsplit}{{Duplex-Split}\xspace}

\newcommand{\mixtral}{{Mixtral}\xspace}

\newcommand{\glam}{{GLaM}\xspace}
\newcommand{\grok}{{Grok1}\xspace}
\newcommand{\opt}{{OPT}\xspace}
\newcommand{\llama}{{Llama3}\xspace}

\newcommand{\eg}{\textit{e.g.}\xspace}

\newcommand{\Numex}{$N_{ex}$\xspace}
\newcommand{\deggrp}{$deg_{grp}$\xspace}
\newcommand{\topk}{top-$k$\xspace}
\newcommand{\lin}{$L_{in}$\xspace}
\newcommand{\lout}{$L_{out}$\xspace}
\newcommand{\nhead}{$N_{head}$\xspace}

\newcolumntype{M}[1]{>{\centering\arraybackslash}m{#1}}

\title{Duplex: A Device for Large Language Models with Mixture of Experts, Grouped Query Attention, and Continuous Batching} 

\usepackage{lipsum}
\usepackage{enumitem}
\usepackage{xspace}
\usepackage{hhline}
\usepackage{makecell}
\usepackage{multirow}
\usepackage[utf8]{inputenc}
\usepackage{graphicx}
\usepackage{lipsum}
\usepackage{blindtext}
\usepackage{tikz}
\usepackage{subfigure}
\usepackage{enumitem}
\usepackage{color}
\usepackage{soul}
\usepackage{comment}
\usepackage{balance}
\usepackage{xfrac}
\usepackage{dblfloatfix}
\usepackage{mathabx}
\usepackage{amsmath}
\usepackage{dashrule}

\usepackage[all]{nowidow}

\author{\IEEEauthorblockN{Sungmin Yun\IEEEauthorrefmark{2},
Kwanhee Kyung\IEEEauthorrefmark{2},
Juhwan Cho\IEEEauthorrefmark{2}, 
Jaewan Choi\IEEEauthorrefmark{2},
Jongmin Kim\IEEEauthorrefmark{2},\\
Byeongho Kim\IEEEauthorrefmark{3},
Sukhan Lee\IEEEauthorrefmark{3},
Kyomin Sohn\IEEEauthorrefmark{3},
Jung Ho Ahn\IEEEauthorrefmark{2}}
\IEEEauthorblockA{\IEEEauthorrefmark{2}\textit{Seoul National University, Seoul, South Korea}, \IEEEauthorrefmark{3}\textit{Samsung Electronics, Hwasung, South Korea}\\
\textit{\{sungmin.yun, kwanhee5, jfcho2, cjw9202, jongmin.kim, gajh\}@snu.ac.kr}\\ \textit{\{bh1122.kim, sh1026.lee, kyomin.sohn\}@samsung.com}}
}

\begin{document}
\maketitle
\pagestyle{plain}


\begin{abstract}
Large language models (LLMs) have emerged due to their capability to generate high-quality content across diverse contexts.
To reduce their explosively increasing demands for computing resources, a mixture of experts (MoE) has emerged.
The MoE layer enables exploiting a huge number of parameters with less computation.
Applying state-of-the-art continuous batching increases throughput; however, it leads to frequent DRAM access in the MoE and attention layers.
We observe that conventional computing devices have limitations when processing the MoE and attention layers, which dominate the total execution time and exhibit low arithmetic intensity (\opb).
Processing MoE layers only with devices targeting low-\opb such as processing-in-memory (PIM) architectures is challenging due to the fluctuating \opb in the MoE layer caused by continuous batching.

To address these challenges, we propose \duplex, which comprises \highp tailored for high-\opb and \lowp to effectively perform low-\opb operation within a single device. 
\duplex selects the most suitable processor based on the \opb of each layer within LLMs.
As the \opb of the MoE layer is at least 1 and that of the attention layer has a value of 4--8 for grouped query attention, prior PIM architectures are not efficient, which place processing units inside DRAM dies and only target extremely low-\opb (under one) operations.
Based on recent trends, \lowp adds more through-silicon vias (TSVs) to enable high-bandwidth communication between the DRAM die and the logic die and place powerful processing units on the logic die, which is best suited for handling low-\opb operations ranging from few to a few dozens.
To maximally utilize the \highp and \lowp, we propose expert and attention co-processing.
By exploiting proper processing units for MoE and attention layers, \duplex shows up to 2.67$\times$ higher throughput and consumes 42.0\% less energy compared to GPU systems for LLM inference.

\end{abstract}


\section{Introduction}
\label{sec:introduction}

Large language models (LLMs) such as GPT-3~\cite{neurips_2020_gpt-3} and GPT-4~\cite{gpt4-techreport} are gaining enormous prominence due to their excellent output quality in a vast range of applications.
Increasing the number of model parameters has been a principal strategy to improve the quality of an LLM under the premise that the larger the model is, the better the output quality becomes.
However, such a straightforward scaling approach cannot persist due to the computational challenge.

Recently, LLMs have employed a Mixture-of-Experts (MoE)~\cite{jmlr-2022-switchtransformer, emnlp-2022-metamoe, icml-2022-DSMoE} construction on top of existing models as an efficient method to increase the number of model parameters without severely bloating the computational overhead, enabling faster training time than LLM without MoE.
The MoE layer replaces the conventional feed-forward network (FFN), mainly composed of matrix-matrix multiplication (GEMM) operations between inputs and trained weights, with a collection of multiple expert FFNs (experts in short) and a gate to choose between experts.
In \glam~\cite{icml-2022-GLaM}, for example, by utilizing only two out of the total 64 experts, the computational load remains equivalent to a 10B model, achieving the effect of using a 143B model.

Adopting continuous batching~\cite{osdi-2022-orca} in inferring LLMs with MoE layers introduces new computational challenges.
Continuous batching is a state-of-the-art scheduling technique that decomposes each inference request into multiple stages and batches these requests at the stage level in a lockstep manner. 
It maximizes the utilization of computing resources through fine-grained scheduling, allowing newly arrived requests to participate with minimal queuing delay, thereby increasing throughput. 
However, this batching strategy also increases the number of experts concurrently used in MoE layers, increasing off-chip memory (DRAM) access for loading the parameters of these experts (e.g., 64 for GLaM).
Further, the amount of DRAM access in the attention layers in conventional LLMs also increases when batching is applied~\cite{asplos-2024-attacc}. 
We observed that the MoE and attention layers, which have low arithmetic intensity (\opb) in most stages, take most of the inference time on conventional systems such as GPUs, falling utilization of its computing resources under 10\%.

\begin{figure*}[!tb]
  \center
\includegraphics[width=0.76\textwidth]{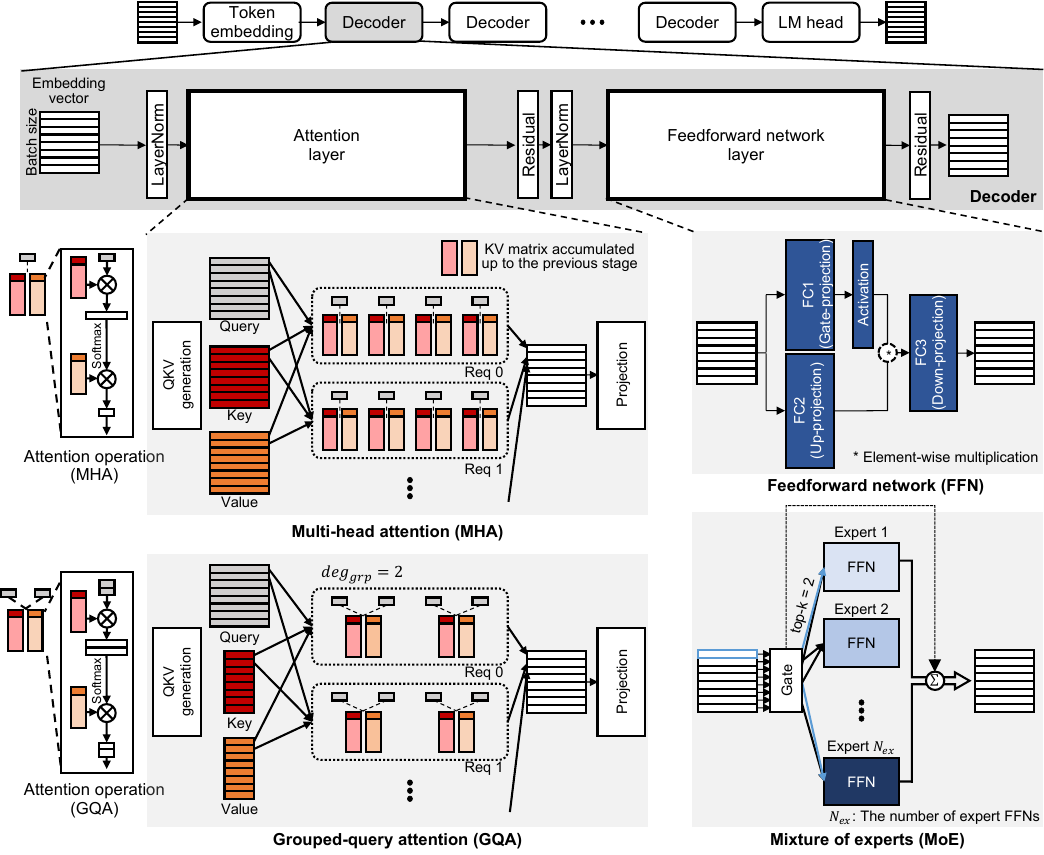}
  \caption{LLM architecture and inference process in a gen stage with batched requests. Attention and FFN layers compose a conventional LLM, whereas GQA and MoE are used in place of these two layers, respectively.
  }
  \vspace{-0.12in}
  \label{fig:llm_model}
  \vspace{-0.09in}
\end{figure*}

However, processing an MoE layer only with devices targeting low-\opb cannot handle the fluctuation of \opb in MoE layers by continuous batching.
Even if the \opb of MoE layers stays at low-\opb in most stages, the arrival of a new request significantly increases the \opb of an MoE layer due to a large number of tokens corresponding to the input sequence passing through the MoE layer.
Processing these layers solely with devices targeting low-\opb struggle due to insufficient computing power, thereby exacerbating the latency of the MoE layer and deteriorating user experience.
In such cases, using devices with high computing power (e.g., GPUs) is preferred.

A recent work~\cite{asplos-2024-attacc} proposed a heterogeneous system that uses processors targeting low-\opb to alleviate DRAM bandwidth bottleneck in the attention layer and utilizes GPUs to process other layers.
Replicating MoE layers to both devices and selecting the devices to process the MoE layers based on the \opb values may handle the fluctuation of \opb in MoE layers.
However, duplicating the parameters of MoE layers, which account for the majority of the model parameters, is inefficient as it requires more devices due to limited memory capacity.
Such systems also restrict the number of batched requests due to the wasted memory capacity than homogeneous systems, consequently reducing the system throughput.

To address the above challenges in the heterogeneous system, we propose \duplex, which integrates \highp, processing units for high-\opb (e.g., GPU), with \lowp that effectively handles low-\opb operations in LLM inferences.
Prior works~\cite{isca-2021-hbm-pim, isscc-2022-aim} have suggested processing-in-memory (PIM) for handling extremely low-\opb (around 1) operations.
As an MoE layer exhibits \opb greater than one, PIM struggles due to its limited computing power.
Further, the introduction of grouped-query attention (GQA), employed in recent models like Llama 2~\cite{arxiv-2023-llama2} and Mixtral~\cite{arxiv-2024-mixtral}, raises the \opb of attention layers to 4 to 8, making PIM that embeds most ALUs in DRAM dies inefficient for processing the attention layers. 
The recent trends of decreasing through-silicon via (TSV) pitches to 22um~\cite{SKhynix-tsv-pitch} and implementing a logic die with logic process~\cite{imw-2017-hbm-dram-technology, todaes-2021-nmp-cnn-hbm} make us propose \lowp, adding more TSVs internally to the high bandwidth memory (HBM) and incorporating processors on the logic die to exploit increased internal bandwidth with massive processing units.

To enhance the utilization rate of \highp and \lowp, we propose expert and attention co-processing for MoE and attention layers.
Going further from processing each layer by one of the \highp and \lowp, this co-processing allows the xPU and Logic-PIM to do finer-grained processing within the MoE layer and attention layer.
We propose expert co-processing that processes the experts utilizing both \highp and \lowp by leveraging the fact that each expert may handle a varying number of tokens and by choosing which processing units to process each expert in the MoE layers.
We assign experts that process a relatively large number of tokens to \highp and the remaining experts to \lowp, reducing the execution time of the MoE layers.
We reduce the execution time of the attention layer using attention co-processing that assigns the attention of each request to both processing units, by assigning high-\opb attention operations of an input request to \highp, and low-\opb ones of ongoing requests to \lowp.

By flexibly utilizing proper processing units for each layer based on the \opb and co-processing optimization, \duplex shows up to 2.67$\times$ higher throughput, 2.57$\times$ lower end-to-end latency, and 42.03\% less energy consumption for LLM inference compared to the baseline GPU without \lowp.

The key contributions of this paper are as follows:
\begin{itemize}[leftmargin=*,nolistsep]
\item Through a detailed analysis, we observe that GPU systems exhibit fundamental limitations in LLM inference due to their lack of support for low-\opb operations.

\item We design \duplex to accelerate MoE and attention layers by housing \highp and \lowp in a single device and selecting the processing units based on the \opb of each layer.

\item We propose \lowp, which incorporates dedicated TSVs connecting the DRAM dies to the logic die where processing units are placed, offering enhanced support for low-\opb (1--32) operations compared to previous PIM architectures.

\item We propose expert and attention co-processing to increase the utilization of \highp and \lowp.
\end{itemize}

\section{Background}
\label{sec:background}

\begin{figure}[!tb]
  \center
\includegraphics[width=1.0\columnwidth]{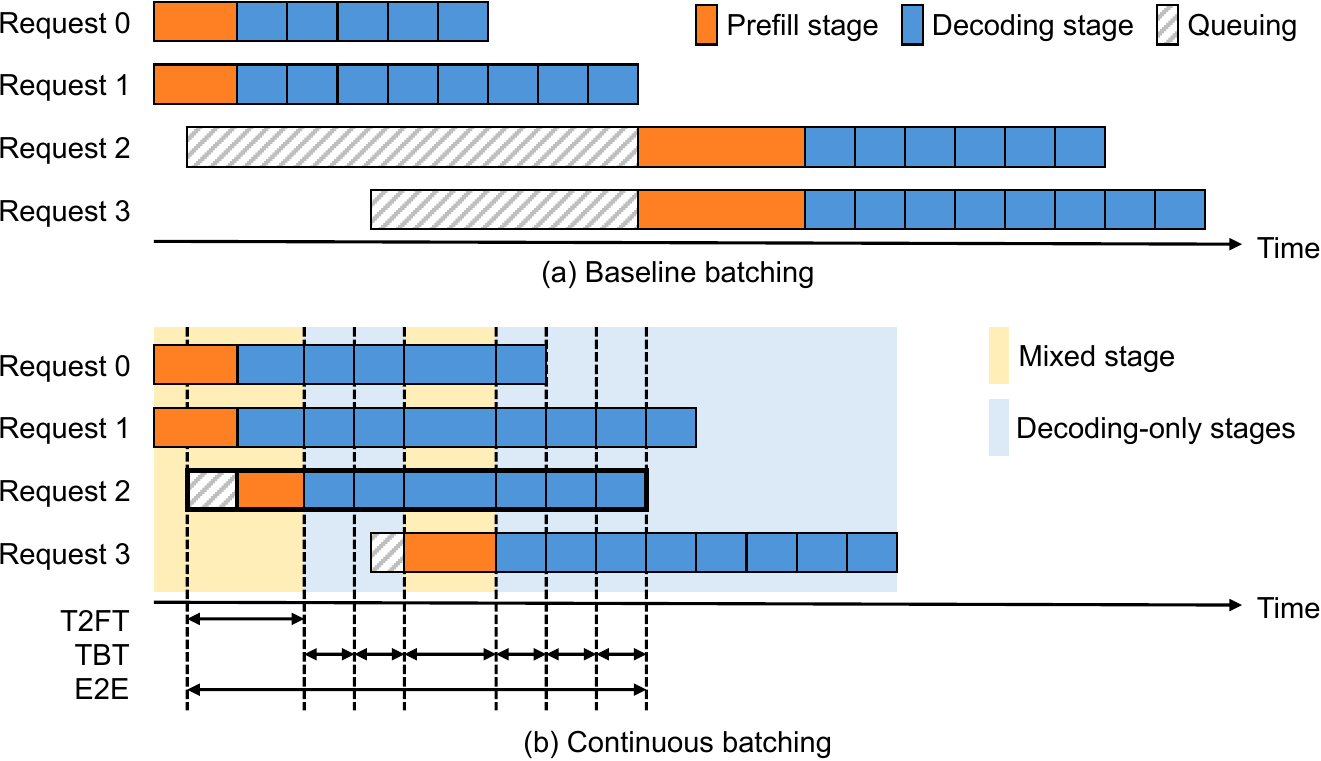}
  \caption{
  (a) Baseline batching, which performs inference at the request level. (b) Continuous batching, which performs inference at the stage level. \ttft, \tbt, and \ete latency values for request 2 are also detailed.
  }
  \label{fig:scheduling}
  \vspace{-0.05in}
\end{figure}

\subsection{Structure of Large Language Models (LLMs)}
\label{sec:background:structure}
Most recently proposed LLMs consist of decoders originally introduced for the transformer model~\cite{neurips_2017_transformer}.
An LLM features sequentially stacked (connected) decoder blocks with a token embedding layer at the beginning and a language modeling (LM) head layer at the end (see Fig.~\ref{fig:llm_model}), where a token is the unit of interpretation.
Upon receiving an inference request expressed as a sequence of tokens, each token is first transformed into a hidden vector with dimensions ranging from thousands to tens of thousands by the token embedding layer.
Hereafter, we refer to the hidden vectors as tokens.
The following decoder block comprises a multi-head attention (MHA) layer, a feed-forward network (FFN) layer, and several other layers that are relatively lightweight from a computational perspective.

MHA first generates query (Q), key (K), and value (V) matrices or vectors from the input sequence through fully-connected (FC) operations.
Each of Q, K, and V is partitioned into evenly sized slices, distributed to \nhead heads (Q\textsubscript{i}, K\textsubscript{i}, and V\textsubscript{i} for i = 0, 1, ..., \nhead$-1$), in which an attention operation is performed.
For each head, a Q slice is multiplied with a K slice, softmax is applied, and the result is multiplied with a V slice.
Finally, projection, another FC layer, is performed for the concatenated results from the heads.
The following FFN layer consists of three FC layers with a gated activation operation (\eg, SiLU~\cite{arxiv-2023-llama2, arxiv-2024-mixtral}) in between.
We assume the use of FP16 format for the data vectors and weights in LLM inference, following conventional practices~\cite{openai-fp16, arxiv-2022-opt}.

\subsection{Mixture of Experts and Grouped-Query Attention}
\label{sec:background:moe_gqa}

To further scale the size of LLMs, hyperscalers, such as OpenAI, Google, Meta, and Microsoft, have employed model variants such as mixture of experts (MoE)~\cite{jmlr-2022-switchtransformer, emnlp-2022-metamoe, icml-2022-DSMoE, arxiv-2024-mixtral, grok1} and grouped-query attention (GQA)~\cite{emnlp-2023-gqa, arxiv-2023-llama2, arxiv-2024-mixtral, grok1}.

MoE improves the output quality while suppressing the increase in the number of operations.
MoE places multiple instances (called experts) of an FFN layer, which already accounts for the majority (about \sfrac{2}{3}) of parameters in non-MoE LLMs.
We denote the number of experts in an MoE layer as \Numex.
A gate placed at the front assigns tokens to different \topk experts; the selected experts are determined based on the input token.
Each token passes through \topk experts, and the final result of the MoE layer is obtained by the weighted summation of each \topk expert's output. 
The memory capacity requirement substantially increases to store the \Numex expert FFNs, and the bandwidth requirement also increases to access the parameters for the selected expert FFNs. 
However, the number of operations is almost the same or slightly larger than that of non-MoE LLMs, as each token only passes through \topk experts, not all experts.

GQA allows heads within an attention layer to form groups, each sharing K and V slices.
Then, Q slices for the heads in the group can perform attention with a form of general matrix-matrix multiplication (GEMM).
The total size of K and V decreases by a factor of \deggrp, the number of heads in a group, alleviating the memory bandwidth bottleneck.
At its extreme, all the heads in a layer can form a single group, which is referred to as multi-query attention (MQA); however, MQA is known to degrade the output quality compared to GQA~\cite{emnlp-2023-gqa, arxiv-2023-llama2}.
Hence, we do not consider MQA in this paper.

\begin{figure*}[!tb]
  \center
\includegraphics[width=0.70\textwidth]{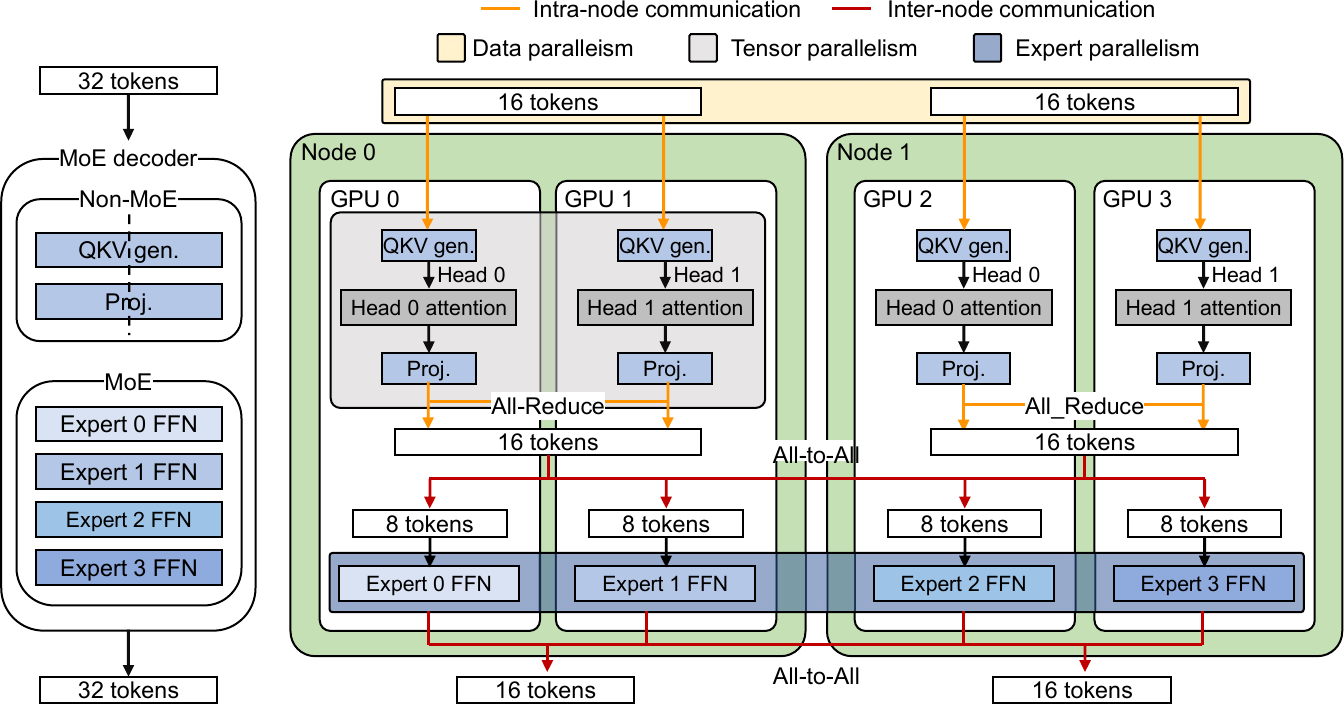}
  \vspace{-0.04in}
  \caption{
  Model distribution methodology and operation flow of an LLM in a multi-node/multi-GPU system~\cite{icml-2022-DSMoE}. For non-expert weights, systems exploit tensor parallelism in the node, and data parallelism across nodes. For expert FFNs, the system allocates each expert FFN to a different GPU.
  }
  \label{fig:gpu_distribution}
  \vspace{-0.12in}
\end{figure*}

\subsection{LLM Inference with Continuous Batching}
\label{sec:background:inference}

LLM inference involves a single prefill (summarization) stage followed by multiple decoding (generation) stages.
The former takes the entire input tokens of length $L_{in}$, passes them through the model, and generates key and value (KV) matrices as well as the first output token.
A sequence of decoding stages follows iteratively; a decoding stage receives a single output token from the previous stage and passes it through the model in a sequential manner.
Each decoding stage generates KV vectors for the input token, which are concatenated to the KV matrices, and a new output token.
Multiple decoding stages are required to constitute the output response of length $L_{out}$.

LLM inference can process multiple requests in a batch to increase serving throughput.
Both on the prefill and decoding stages, the FC layers from QKV generation, projection, and FFN layers can be batched to form GEMM operations between the batched input tokens and the weight matrices.
However, batching requests is not effective for the attention operation.
The attention operation must be performed separately for each request because the unique KV matrices corresponding to the context of each request are used.

To increase serving throughput, continuous batching~\cite{osdi-2022-orca} is widely used.
It divides each LLM inference into multiple stages and batches the requests at the stage level (see Fig.~\ref{fig:scheduling}(b)).
This stage-level scheduling can reduce the queuing delay of new requests and the time-to-first token (\ttft), the latency it takes for the first token to be generated upon request arrival.
We categorize each stage into the following two types depending on the presence or absence of a prefill stage request.
1) \mixedstage: prefill stages of newly added requests are batched with decoding stages of existing requests. 2) \deconlystage: all requests of a batch are in the decoding stage if there is no new request to be served at the moment a new stage starts.
We refer to the latency between two consecutive token generations as token-between-token latency (\tbt) and the request handling latency from arrival to completion as end-to-end latency (\ete).
In each stage, we refer to the requests performing decoding as decoding sequences and those performing prefill as prefill sequences.
Hereafter, batch size is determined by the number of requests in a stage.

\begin{figure*}[!tb]
  \center
\includegraphics[width=0.95\textwidth]{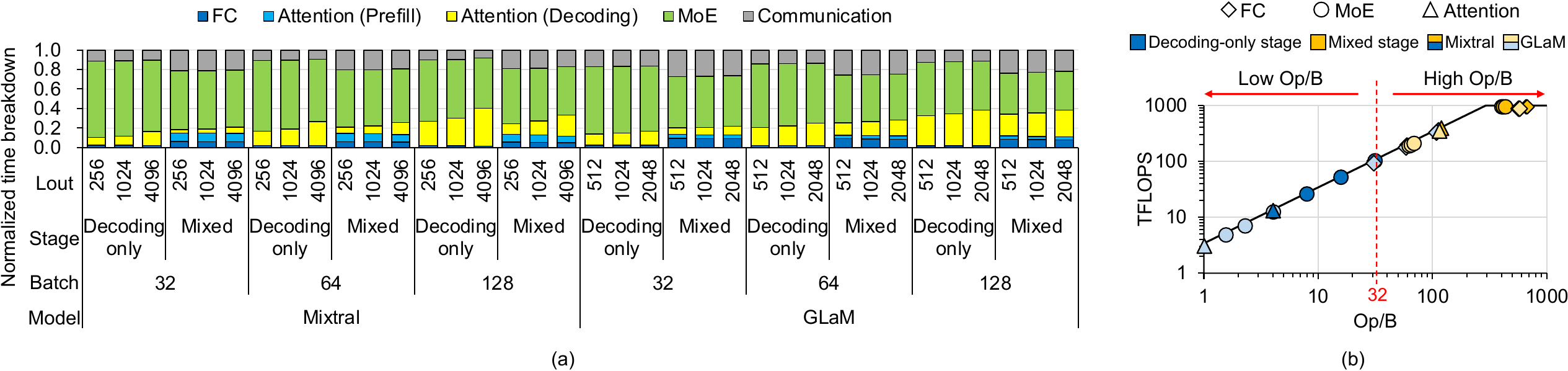}
  \vspace{-0.1in}
  \caption{
  (a) Execution time ratio of each operation in Mixtral~\cite{arxiv-2024-mixtral} and GLaM~\cite{icml-2022-GLaM} varying \lout and batch size while \lin = 2048. Mixtral (GLaM) uses \deggrp$=4$ $(1)$ for the attention layer and uses 8 (64) experts in the MoE layer with each token selecting the top-2 experts. (b) The roofline graph for each model on GPUs with varying batch sizes (32--128) when \lin = 2048 and \lout = 1024. Details of systems are in Section~\ref{sec:experimental_setup}.
  }
  \label{fig:roofline}
  \vspace{-0.15in}
\end{figure*}

\subsection{High Bandwidth Memory (HBM)}
\label{sec:background:hbm}
HBM has a 3D-stacked structure with one logic die at the bottom and multiple DRAM dies.
The logic die consists of I/O circuitry, memory built-in-self-test, and testing and debugging units~\cite{isca-2021-hbm-pim, pact-2018-3d-xpath}.
Through silicon vias (TSVs) connect the DRAM dies to the logic die.
We focus on 8-hi HBM3, deployed on the latest GPUs (e.g., NVIDIA H100).
In HBM3, four DRAM dies form a rank, and each DRAM die has eight pseudo channels. Each pseudo channel is connected to four bank groups of four banks, totalling 16 banks in the rank. 
The banks share external wires within a single pseudo channel, allowing them to read data from only one bank at a time.

\section{Computational Analysis}
\label{sec:motivaiton}

We analyze how MoE-based LLMs with MHA or GQA perform on a multi-GPU system and explore available options to enhance the performance.
We follow the data/model/expert parallelism methodologies from \cite{icml-2022-DSMoE} for the job distribution among the GPUs.
Fig.~\ref{fig:gpu_distribution} shows an exemplar model distribution of an LLM with four expert FFNs in the system consisting of two nodes with two GPUs each.
The system uses expert parallelism for the MoE layers, which distributes expert FFNs across the GPUs.\footnote{If the number of GPUs ($N_{\text{GPUs}}$) exceeds the number of experts, then each expert is allocated \(\frac{N_{\text{GPUs}}}{N_{\text{ex}}}\) GPUs using tensor parallelism.}
For the FC layers excluding MoE, the system uses tensor parallelism by partitioning rows or columns of a weight matrix within a node and data parallelism by distributing requests across the nodes.

\subsection{Computational Analysis of MoE and Attention Layers}
\label{sec:motivation:compute_analysis}

The MoE and attention layers are dominant in both the \deconlystage and the \mixedstage (Fig.~\ref{fig:roofline}(a)).
Although adopting MoE increases the amount of computation just by $k$, the number of expert FFNs chosen by a gate (\eg, $k=2$~\cite{icml-2022-GLaM, arxiv-2024-mixtral, grok1}), independent requests as a whole are expected to explore most of the expert FFNs in the model (\Numex);
thus, memory access skyrockets to load \Numex expert FFNs, which in turn raises latency.
In the case of the attention layer, as shown in prior work~\cite{asplos-2024-attacc}, the throughput improvement from batching diminishes because each request accompanies its own KV matrices.
Therefore, as the sequence length and the batch size increase, the significance of attention layers increases.

MoE and attention layers exhibit low \opb in the \deconlystage (see Fig.~\ref{fig:roofline}(b)), which severely reduces GPU utilization.
Compute utilization becomes lower than 11\% for the MoE layer and 2.06\% for the attention layer on GPUs.
Because the tokens in the batch are distributed among the experts by gate, each expert processes a relatively small number of tokens because $k$ is smaller than \Numex.
Still, multiple requests can share the same expert, resulting in the \opb becoming higher than one.
Second, as multiple heads share KV matrices, the attention layer exhibits higher \opb for GQA (Mixtral) than for MHA (GLaM), but \opb remains low even with GQA.
Unique KV matrices exist for each request and for each head (MHA) or each group (GQA), resulting in a GEMV with a Q vector or a GEMM with a narrow \deggrp-wide Q matrix, where \deggrp ranges from four to eight~\cite{arxiv-2024-mixtral, grok1, arxiv-2023-llama2}.
This computation exhibits low \opb.

\begin{figure*}[!tb]
  \center
\includegraphics[width=0.99\textwidth]{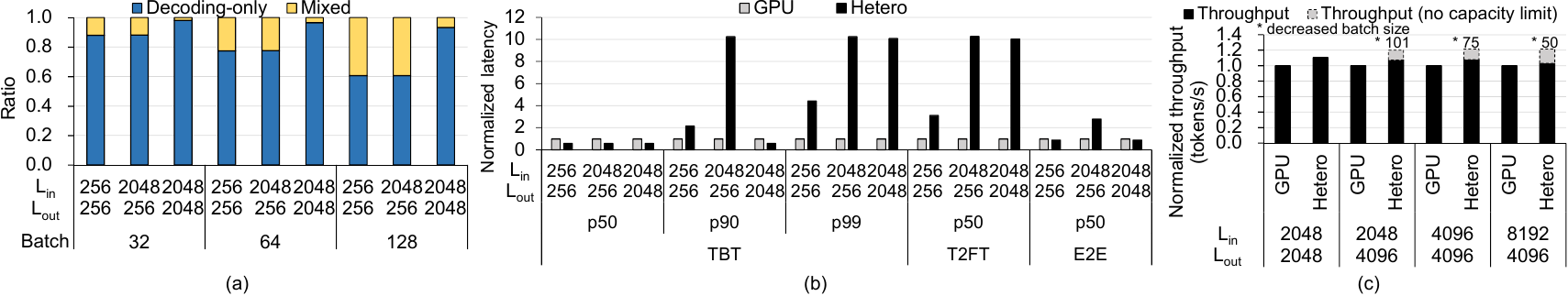}
  \vspace{-0.1in}
  \caption{(a) The ratio of \deconlystage to \mixedstage in \mixtral on a GPU system. (b) The normalized latency of a heterogeneous system compared to a \gpu system in \mixtral with a batch size of 32. The \gpu system consists of four GPUs, while the heterogeneous system consists of two GPUs and two \lowp{s} (details in Section~\ref{sec:duplex}). (c) The normalized throughput of the heterogeneous system over the \gpu system in \mixtral with a batch size of 128.
  }
  \label{fig:latency_cdf}
  \vspace{-0.1in}
\end{figure*}

The request batch size also significantly impacts the \opb except for the attention layer, which is performed separately for each request.
In particular, utilizing a larger batch size increases the \opb of the MoE layer as more requests in a batch share the same expert.
Nevertheless, for practical batch sizes abiding by the latency limitation imposed by the service level objective (SLO) and memory capacity for KV matrices, the attention and MoE layers stay in the low \opb region.

In the case of \mixedstage, the \opb of MoE and attention layers increases.
A new request added to \mixedstage increases the number of tokens that select each expert, increasing the \opb of the MoE layer.
The new request causes numerous tokens, much more than \deconlystage by $L_{in}$, pass through the MoE layer, resulting in a higher number of tokens processed per expert.
In the attention layer, the operation for the new request exhibits a high \opb because $L_{in}$ Q slices share the same KV matrices for each head.

\subsection{Limitations of Heterogeneous Systems}
\label{sec:motivation:hetero_limitations}
We discover that most stages in LLM inference with continuous batching are \deconlystage{s}.
This is because each request consists of a single prefill stage and multiple ($L_{out}$) decoding stages.
Fig.~\ref{fig:latency_cdf}(a) shows the ratio of decode-only and mixed stages in Mixtral for various $L_{in}$ and $L_{out}$ values in a four-GPU system (detailed in Section~\ref{sec:experimental_setup}).
It can be observed that the \deconlystage is dominant for all cases.
Thus, it is important to quickly process the \deconlystage to reduce the \ete latency and improve serving throughput.

To accelerate the \deconlystage, it is necessary to speed up the MoE layers and attention operations, which occupy most of the time in the \deconlystage and have low-\opb characteristics. Hence, one can design a heterogeneous system (hetero system)~\cite{asplos-2024-attacc} that includes device nodes with high memory bandwidth and low computing power dedicated to MoE and attention layers alongside conventional GPUs.
We suppose such a hetero system consisting of two GPUs and two \lowp(detailed in Section~\ref{sec:duplex}) and compare it with a four-GPU system.
We assume that the \lowp processes all MoE layers of all stages and the attention operations of \deconlystage{s} to avoid MoE weight duplication.
This hetero system can reduce the median (p50) \tbt and \ete latency as shown in Fig.~\ref{fig:latency_cdf}(b).

However, the 90th (p90) and 99th (p99) percentile \tbt and median \ttft latency values show significant increases due to the limited computing power of the devices targeting low-\opb operations.
As the MoE layers exhibit high 
\opb in the \mixedstage, devices with high memory bandwidth but low computing power become severely compute-bound for these layers. 
As \lin increases, the higher \opb in the MoE layer leads to increased \ttft and tail latency in \tbt, and even in \ete for some cases. 
\tbt and \ttft are critical performance metrics for LLM inference, especially in conversational tasks~\cite{isca-2024-splitwise}, which often involve multiple rounds of dialogues between the user and the chatbot.
Because each round is processed as a separate request, \lin continues to increase as the conversation progresses, exacerbating the problem.

To prevent the tail latency from skyrocketing, it is necessary to duplicate the weights of the MoE layers on GPUs with high computing power and process the MoE layer of the \mixedstage on these GPUs.
However, duplicating the weight of MoE layers, which take the most model weights, is highly inefficient, requiring more devices due to limited memory capacity.
Further, such hetero systems limit memory capacity for KV matrices, which increases proportionally to batch size, than homogeneous systems, thereby reducing the maximum batch size and thus hindering system throughput (see Fig.~\ref{fig:latency_cdf}(c)).

\section{Duplex: Devices for Efficient LLM Inference}
\label{sec:duplex}
To address the challenges in the hetero system, we propose \duplex, which configures separate processing units for high-\opb and low-\opb operations that share device memories.
Appropriate processing units are selected for each operation based on the stage.
The low-\opb unit handles the MoE layers during the \deconlystage as well as the attention layers of the \deconlystage and of decoding sequences in the \mixedstage.
The high-\opb unit manages the rest.
We opt for an HBM-based system to provide high memory bandwidth.

\subsection{Implementation of High Op/B Processors}
\label{sec:duplex_implement_high_opb}

For high-\opb operations, conventional accelerators, such as GPUs and TPUs, are eligible candidates for a high \opb processor. 
We assume that a popular GPU architecture equipped with HBM serves as a high \opb processor for \duplex.
Hereafter, we refer to the processor as \highp.
Numerous processing units in \highp provide extremely high computational throughput, but the HBM bandwidth is limited due to the physical limitations of the interposer connecting HBM stacks with the main computing die.

\subsection{Implementation of Low Op/B Processors}

\label{sec:duplex_implement_low_opb}

\begin{figure}[!tb]
  \center
\includegraphics[width=1.0\columnwidth]{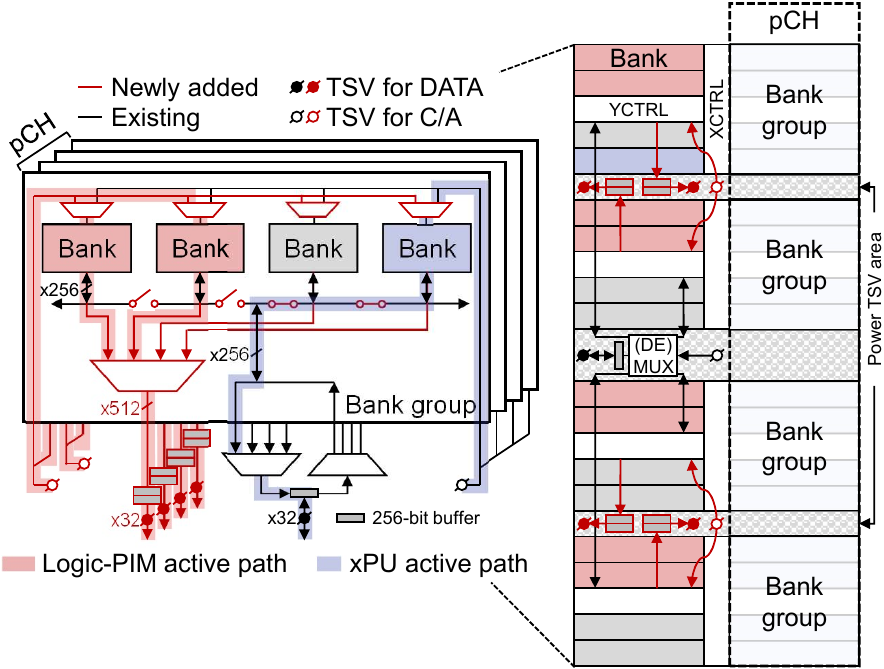}
  \vspace{-0.1in}
  \caption{Our DRAM die microarchitecture of HBM3 for \lowp. \lowp and xPU can operate simultaneously through independent active paths.}
  \label{fig:duplex_microarchitecture}
  \vspace{-0.1in}
\end{figure}

\begin{figure*}[!tb]
  \center
\includegraphics[width=0.95\textwidth]{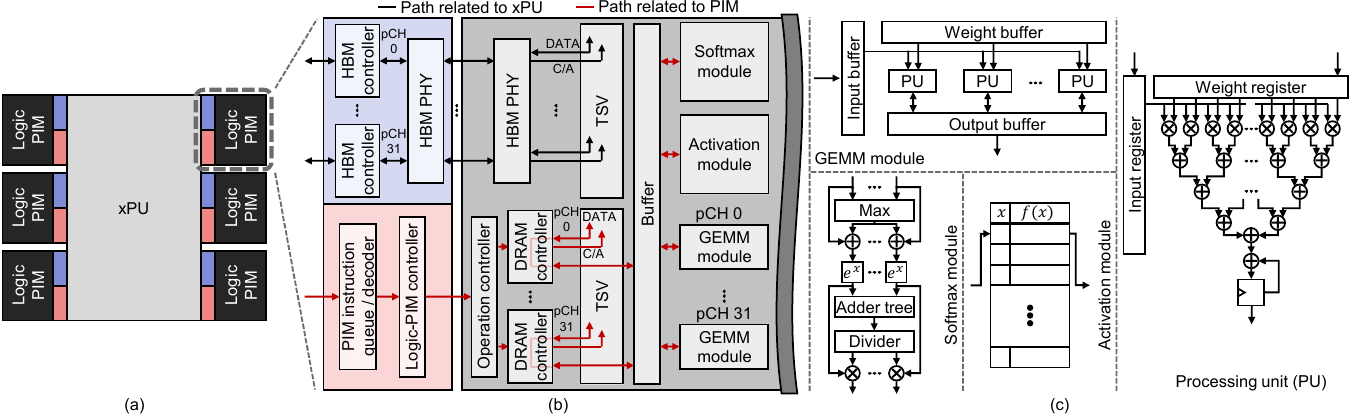}
  \vspace{-0.08in}
  \caption{(a) Top view of Duplex chip, (b) the architecture of Logic-PIM, and (c) the details of processing units.
  }
  \label{fig:duplex_chipdesign}
  \vspace{-0.1in}
\end{figure*}

The MoE and attention layers, which exhibit low \opb ranging from 1 to 32, face challenges when performed using the PIM architectures in prior studies~\cite{isca-2021-hbm-pim, isscc-2022-aim, asplos-2024-attacc, asplos-2024-neupim}, which place processing units inside DRAM dies. 
The banks share external wires within a single pseudo channel, allowing them to read data from only one bank at a time. 
In contrast, a PIM architecture places an internal computing unit in each bank, where all those can read data from its bank simultaneously, improving effective memory bandwidth.
While such PIM architectures are effective for extremely low-\opb operations (under 1) by exploiting the high internal bandwidth of DRAM dies, their performance is suboptimal for operations in the MoE and attention (GQA) layers. 
Populating more processing units with these PIM architectures~\cite{isca-2021-hbm-pim, isscc-2022-aim} is not cost-effective due to the significant area overhead involved in integrating processing units with DRAM processing technologies.
In commercially available PIM devices such as~\cite{isca-2021-hbm-pim, isscc-2022-aim}, even when the compute-to-memory-bandwidth ratio is one, processing units already occupy 27\% and between 20\% and 25\% of the DRAM die area, respectively.

Considering the required range of \opb (1 to 32), we design the target architecture for low-\opb units to achieve 4$\times$ the memory bandwidth of conventional HBM, with a compute-to-memory-bandwidth ratio of eight. 
Because the logic die can leverage the logic process~\cite{imw-2017-hbm-dram-technology, todaes-2021-nmp-cnn-hbm}, more processing units can be placed on the logic die than on the DRAM die of the same area.
However, merely adding processing units to the logic die does not offer any memory bandwidth benefits compared to \highp.
We observe \emph{a reduction in the TSV pitch in recent HBMs from 50um to 22um}~\cite{SKhynix-tsv-pitch}, which allows quadrupling the number of TSVs with only a 9\% area overhead (detailed in Section~\ref{sec:eval:area_overhead}).
Motivated by this technology trend, we propose \lowp, which increases the internal bandwidth by adding more TSVs and adds processing units to the logic die which can utilize the logic process.

\subsection{Microarchitecture of \lowp}
In designing \lowp, we set two main objectives of minimizing modifications to DRAM and reducing the internal DRAM datapath length to reduce the energy required to read data~\cite{MICRO-2017-fine-grained-dram}.
In providing 4$\times$ higher memory bandwidth, simply increasing the bandwidth of each bank incurs a significant overhead of quadrupling the prefetch size of the banks and increasing the I/O datapath width by 4$\times$.
This results in a 77\% increase in the size of the DRAM banks and necessitates changes to the DRAM bank layout~\cite{MICRO-2017-fine-grained-dram}.

Instead, we increase the number of banks operating simultaneously without modifying the structure of the DRAM banks.
Conventional memory systems share bank I/O and bank group I/O, allowing data to be read from only one bank at a time. 
We place switches between each bank I/O and separate their paths to enable reading data simultaneously from multiple banks (see Fig.~\ref{fig:duplex_microarchitecture}).
Because reading data from the same bank group takes twice as long (tCCD\_L) as that from different bank groups (tCCD\_S), we simultaneously read from eight banks to achieve 4$\times$ higher bandwidth. 
We divide 16 banks for a single rank and a pseudo channel into upper (colored red in Fig.~\ref{fig:duplex_microarchitecture}) and lower banks and make each group of eight banks operate as one unit, which we refer to as bank bundle.

We integrate additional TSVs for \lowp in the conventional HBM's power TSV area instead of data TSV area. 
To transmit data read from each bank in a \bankbundle to the logic die, we must connect the \lowp-path I/O from each bank group to the TSVs.
Adding TSVs for \lowp to the existing data TSV area would result in longer datapaths from each bank group, thereby consuming more energy~\cite{MICRO-2017-fine-grained-dram}. 
By placing \lowp TSVs near the areas for power TSVs, we reduce the length of \lowp-path I/O from each bank to TSVs and minimize the wiring overhead.

\highp and \lowp can read data simultaneously using \bankbundle parallelism.
\lowp sends the same command/address (C/A) to the target \bankbundle, thus simultaneously reading data from eight banks. \lowp reads a total of 512 bits from two banks per bank group at intervals defined by tCCD\_L, which travels to the I/O buffer in the additional TSVs installed, and the data is transferred through the TSV to the logic die. 
In the case of the \highp path, a simple switch separates it from the \lowp datapath, allowing \highp to read data from the other bank bundles even when \lowp is accessing data.
Each pseudo channel comprises four bank bundles, organized into two ranks with two bank bundles per rank, with indices set from one to four. 
To prevent \bankbundle conflicts when simultaneously using \lowp and \highp, we strategically allocate the weights of models and KV matrices considering the index of bank bundles.

\begin{figure}[!tb]
  \center
\includegraphics[width=0.64\columnwidth]{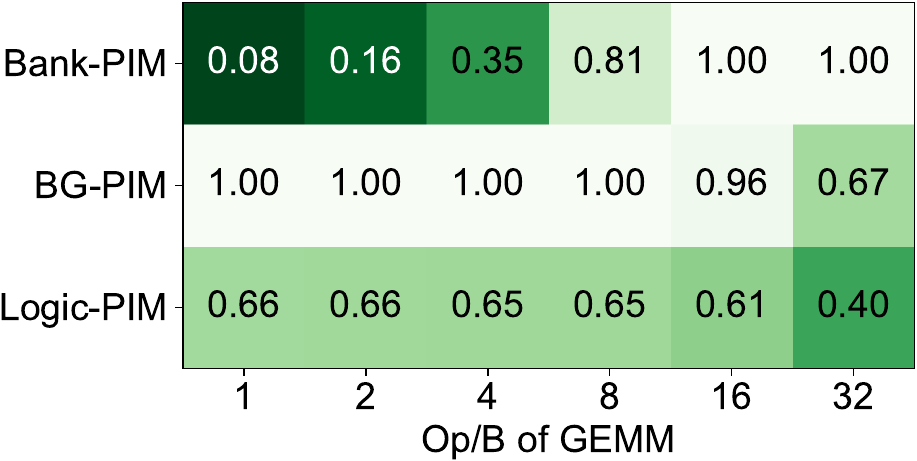}
  \vspace{-0.05in}
  \caption{The normalized energy-delay-area product (EDAP) of Bank-PIM, BankGroup-PIM, and Logic-PIM by \opb of FP16 GEMM operation. Weight matrix is (16384$\times$4096). Details about each architecture are in Section~\ref{sec:experimental_setup}.
  }
  \label{fig:edap}
  \vspace{-0.10in}
\end{figure}

\subsection{\duplex Architecture}

Fig.~\ref{fig:duplex_chipdesign}(a) shows the overall design of \duplex. 
An \highp in the center is responsible for high-\opb operations. 
A \lowp (Fig.~\ref{fig:duplex_chipdesign}(b)) is connected to a conventional HBM controller to read data for high-Op/B operations, and is also connected to a \lowp controller, which controls the processing units in \lowp. 
Unlike HBM controllers, the \lowp controller does not receive data from \lowp.
Instead, all computations are performed on the logic die. 
Thus, only control pins are used to connect \lowp with the controller, resulting in minimal pin overhead.

\lowp consists of a simple DRAM controller for fetching data from HBM, GEMM modules for performing low-\opb GEMM operations, a buffer, and modules for softmax and activation functions. 
The \lowp controller sends \lowp the starting addresses of weights and inputs, as well as the dimensions of the GEMM to the operation controller. Upon receiving a compute request from the \lowp controller, the operation controller fetches data via the DRAM controller and performs computations using the GEMM module.
The activation module handles the activation in the MoE layers, and the softmax module is used for the attention layer.

\subsection{Comparing \duplex with prior PIM architecture}
\duplex is more suited for contemporary LLMs than prior PIM architectures that add processing units to DRAM dies.
\duplex exhibits a better energy-delay-area product (EDAP) for GEMM operations above 8 \opb over prior DRAM die-based PIM architectures (see Fig.~\ref{fig:edap}, prior PIM architectures are detailed in Section~\ref{sec:experimental_setup}).
At \opb under eight, Bank-PIM, which can utilize the highest memory bandwidth, shows the best EDAP. 
However, as the \opb of operations increases, Bank-PIM, with its limited computing power, becomes less efficient compared to \lowp. 
Although BankGroup-PIM has the same memory bandwidth and computing power as \lowp, it always exhibits a higher (worse) EDAP due to having all its processing units and buffers on the DRAM die, resulting in a larger area overhead than \lowp.

\begin{figure}[!tb]
  \center
\includegraphics[width=0.85\columnwidth]{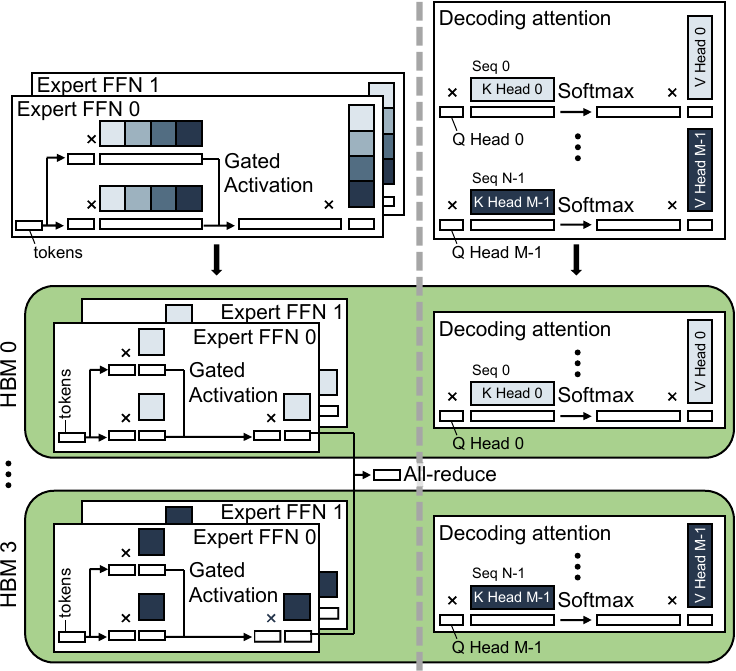}
  \vspace{-0.1in}
  \caption{MoE and attention layers distributed among HBM stacks in a device. The shades of blue indicate which HBM each chunk of data is stored in.}
  \label{fig:duplex_job_distribution}
  \vspace{-0.1in}
\end{figure}

\section{End-to-end LLM inference using Duplex}
\label{sec:duplex_llm_inference}

\subsection{Processing MoE and Attention Layers Using \lowp}
\label{sec:duplex_operation_logic_pim}
We propose a method for distributing computations assigned to each device across \lowp stacks. 
Fig.~\ref{fig:duplex_job_distribution} illustrates how the computations for the MoE and attention layers of decoding sequences are divided into four \lowp stacks. 
For the expert FFNs in an MoE layer, assigning a different expert to each \lowp could lead to varying execution times across \lowp due to differences in the number of tokens processed by each expert. 
We distribute each expert FFN computations across all \lowp for load balancing across \lowp.
However, this approach may necessitate inter-\lowp communication to obtain the final results. 

To minimize communication, we first distribute the weights for gate-projection and up-projection by slicing them column-wise across all \lowp. 
Since gated activation is an element-wise operation, it can be performed without additional data movement. 
Afterward, when computing down-projection, each \lowp ends up holding partial sums of the final output of the expert FFN. 
Performing an all-reduce operation among the \lowp within the device yields the final result of the expert FFN. 
This all-reduce operation is processed by an \highp, which reads and processes all the partial sums stored in the memory of each \lowp.
We minimize communication overhead by conducting a single all-reduce operation after all expert FFNs have completed their computations, rather than performing it after each expert FFN.

\duplex uses request and head parallelism for attention operations. 
Because attention operations among different requests have no data dependency, requests can be fully parallelized. 
As each head operates on separate slices of the Q vector and KV matrices, heads can also be fully parallelized.

\begin{figure}[!tb]
  \center
\includegraphics[width=1.00\columnwidth]{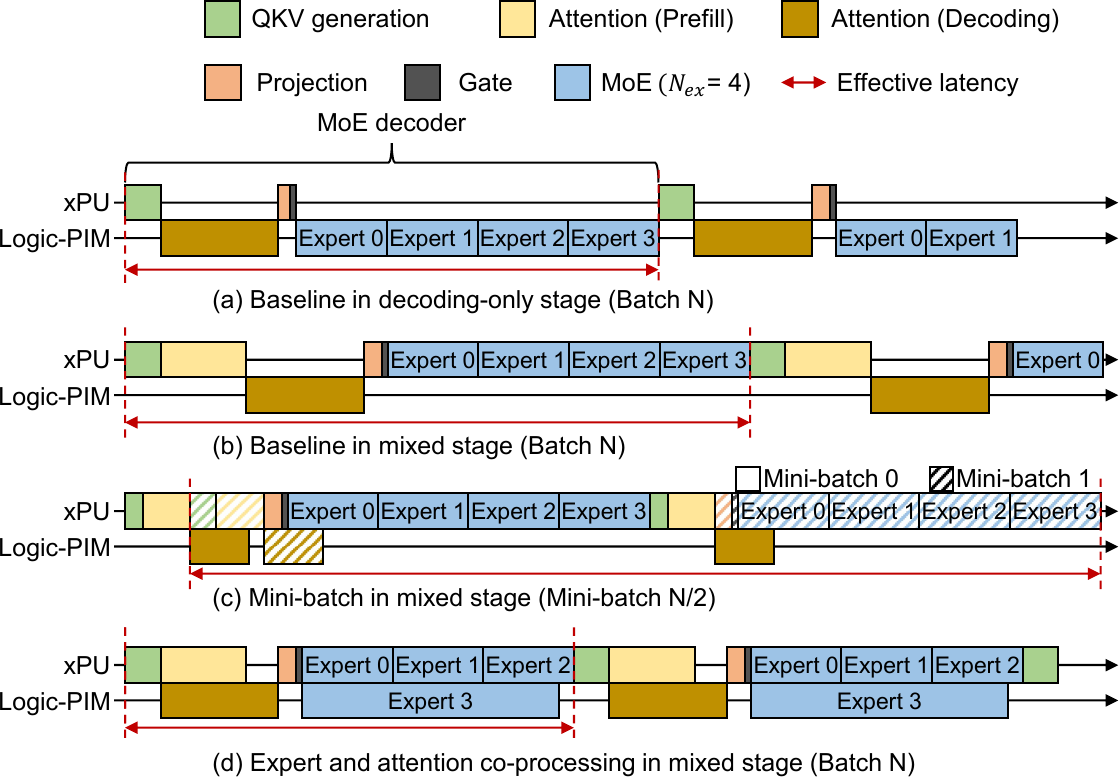}
  \vspace{-0.13in}
  \caption{
  Operation flows of \duplex. Comparison of (a)--(d) for the same total batch size (N) with the same capacity of KV cache, considering the total device memory capacity. For convenience, the \deconlystage is depicted at a larger scale compared to the \mixedstage.
  }
  \label{fig:duplex_operation_flow}
  \vspace{-0.12in}
\end{figure}

\subsection{Expert and Attention Co-processing}
\label{sec:duplex_co-processing}
To increase the utilization of \highp and \lowp, we propose expert and attention co-processing.
In LLM inference, the presence of data dependencies between layers makes simultaneous computation in xPU and Logic-PIM challenging. 
Fig.~\ref{fig:duplex_operation_flow}(a) and (b) illustrates a na\"ive operation flow where only an \highp or \lowp is used at any given time.

One simple way to simultaneously utilize \highp and \lowp is by dividing the workload into two independent mini-batches that have no data dependencies between them (Fig.~\ref{fig:duplex_operation_flow}(c))~\cite{asplos-2024-neupim}. 
These mini-batches enable simultaneous operations in xPU and Logic-PIM by alternating between layers, but it has disadvantages. 
The batching effect is reduced for the FC and MoE layers as these layers operate with half the batch size compared to the baseline method, leading to decreased data reuse. 
While attention operations are unaffected because each request processes independent values, the FC and MoE layers see no reduction in execution time when their operations are memory-bound, even with a reduced batch size. 
This can lead to increased latency compared to the baseline. 
Moreover, processing the same batch size doubles the amount of model parameters being read, resulting in higher DRAM read energy.

We propose expert and attention co-processing to increase the utilization of both processors, while maintaining the batching effect of the FC and MoE layers.
Expert FFNs in MoE layers do not have data dependencies between them, enabling simultaneous computation.
Because all tokens independently pass through a gate to select experts, the number of tokens processed by each expert can vary.
Experts with relatively fewer tokens are processed in \lowp, while the rest are handled in \highp in both \deconlystage and \mixedstage, thus preserving the batching effect of MoE layers while reducing its latency.

Because the number of tokens processed by each expert is determined after passing through gates, selecting which experts to process on each device may incur overhead. 
\duplex must decide which experts to allocate to either \highp or \lowp based on the time to process each expert with how many tokens are processed with each processing unit.
To minimize overhead, \duplex preliminarily estimates and stores the processing times for experts in both \highp and \lowp, depending on the number of processed tokens. 
At runtime, \duplex uses this lookup table to determine which experts to process in \lowp. 
First, \duplex calculates the total time to process all experts using only \highp.
Then, it progressively assigns the experts with the fewest tokens to \lowp, aiming to find the best combination for processing experts.
Then, \highp sends PIM instructions to \lowp to process the corresponding experts.
This lookup table-based decision-making is considerably faster than the actual execution of the expert layer, and its time impact can be considered negligible.

Using expert parallelism may diminish the impact of expert co-processing. 
With fewer experts processed in each device, the degree to which they can be split between xPU and Logic-PIM is limited, reducing the effectiveness of expert co-processing. 
Thus, we choose to apply tensor parallelism for MoE layers, splitting each expert across all devices. 
In the multi-node system, where the bandwidth between nodes is relatively lower than within the same node, we use expert parallelism between nodes and tensor parallelism within nodes.

Second, the mixed stage involves processing attention for both the decoding sequences and the prefilling sequences. 
As attention operations can be processed individually for each request, the attention of prefilling sequences is handled by the \highp, and that of decoding sequences is processed in \lowp, allowing us to process the attention layer more quickly.

\subsection{Memory Allocation and Management}
\label{sec:duplex_memory_space_allocation}
To support co-processing, we divide all the memory space in the device into four sections based on the index of the bank bundle. 
Each memory space uses bank bundles in all channels.
For the expert FFNs, we allocated them one by one across these four memory spaces. 
During expert co-processing, \duplex processes all the experts within the memory space with either \lowp or \highp, thus preventing any bank bundle conflicts between \lowp and \highp.

For the KV cache used in the decoding sequence, we have alternately allocated it among three of the memory spaces, while the remaining memory space is designated for storing Q, K, and V matrices used in the attention of prefilling sequences, thus enabling attention co-processing.
As the K and V matrices used in the prefilling sequences should be cached for the next stages, we need to migrate the K and V matrices to the other bank bundles for the next stage.
After the attention operation is finished, \highp moves the K and V matrices to the bank bundle designated to store KV cache.
Considering that this migration is performed only once, the overhead is negligible.
The parameters for the other layers are used exclusively in \highp and are allocated in any remaining memory spaces.
\section{Experimental Setup}
\label{sec:experimental_setup}

We compare \duplex with a baseline NVIDIA H100 GPU~\cite{nvidia-h100}.
To quantitatively evaluate the performance improvement, we also compare with \doublegpu, a system equipped with twice as many devices.
We configured \highp in \duplex to have the specifications equivalent to H100, which replaced HBM3 with our proposed \lowp with no change in memory capacity (16 GB per stack, 8-hi (two ranks) per stack, and 80 GB per device).
\lowp gains additional 4$\times$ memory bandwidth over conventional HBM3 by adding dedicated TSVs from DRAM dies to a logic die.
We incorporated processing units in \lowp to achieve peak FLOPS for 8 \opb (21.3 TFLOPS per \lowp stack).
For \bankpim, we assume 16$\times$ bandwidth than conventional HBM with a peak \opb of 1, twice as high as HBM-PIM~\cite{isca-2021-hbm-pim}.
\bgpim has the same memory bandwidth and computing power as \lowp, but processing units are in the DRAM die.
Both \bankpim and \bgpim have softmax and activation units on the logic, similar to \lowp.

\begin{table}[!tb]
\vspace{-0.0in}
\caption{Model Configuration Used for Evaluation}
\vspace{-0.2in}
\begin{center}
\resizebox{1\columnwidth}{!}{
\centering
\begin{tabular}{r||c|c|c|c|c|c|c|c}
\toprule 
{Model} & Param. & \# layer & Hidden & Interm. & \# head & \deggrp & \Numex & \topk \\
\midrule
    Mixtral & 47B & 32 & 4096 & 14336 & 32 & 4 (GQA) & 8 & 2 \\
    GLaM & 143B & 32 & 4096 & 16384 & 32 & 1 (MHA) & 64 & 2 \\
    Grok1 & 314B & 64 & 6144 & 32768 & 48 & 6 (GQA) & 8 & 2 \\
\midrule
    OPT & 66B & 64 & 9216 & 36864 & 72 & 1  (MHA) & - & - \\
    Llama3 & 70B & 80 & 8192 & 28672 & 64 & 8 (GQA) & - & - \\ 
\bottomrule 
\end{tabular}
}
\end{center}
\vspace{-0.2in} 
\label{tab:model_configuration} 
\end{table}

\begin{figure*}[!tb]
  \center
\includegraphics[width=1.0\textwidth]{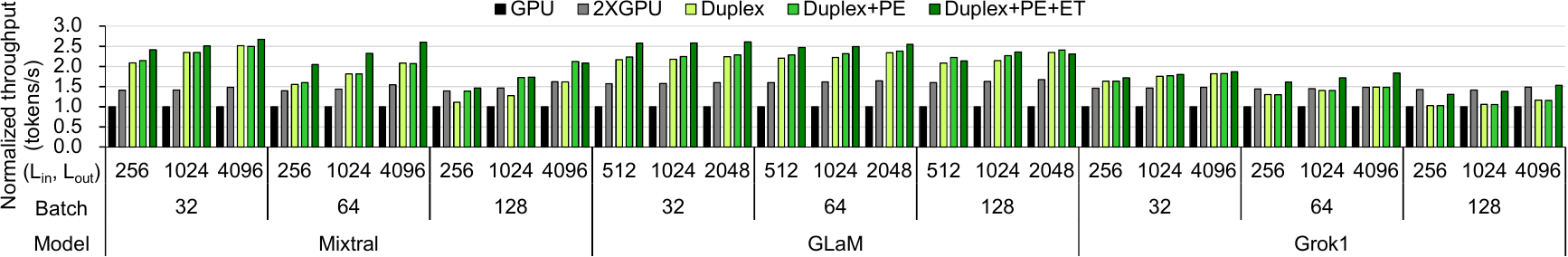}
    \vspace{-0.15in}
  \caption{
  The normalized throughput of \mixtral, \glam, and \grok for various (\lin, \lout) from (256, 256) to (4096, 4096), and batch sizes (32--128).}
  \label{fig:speedup_graph}
  \vspace{-0.1in}
\end{figure*}

\begin{figure}[!tb]
  \center
\includegraphics[width=1.0\columnwidth]{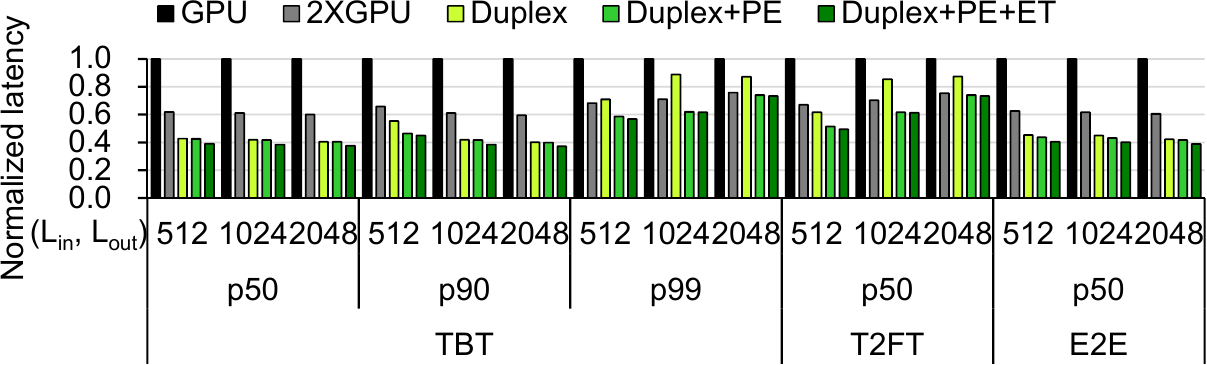}
  \caption{
  The normalized latency (\tbt, \ttft, \ete) of \glam for various (\lin, \lout) from (512, 512) to (2048, 2048) with a batch size of 64.
  }
  \label{fig:eval_latency_bar_graph}
  \vspace{-0.2in}
\end{figure}

To fairly compare the \duplex and GPU, we set the memory capacity of each system to be the same.
With eight or fewer devices, we assume they are interconnected using bidirectional 900GB/s NVLink, similar to an HGX system~\cite{nvidia-hgx}. For configurations with more than eight devices, we assume that each set of eight devices forms a node and that these nodes are interconnected via a system with a 400GB/s Infiniband~\cite{infiniband-whitepaper}.

We developed a cycle-accurate simulator for modeling systems with \duplex and GPUs using Ramulator~\cite{ramulator2-github,cal-2023-ramulator2}.
Our simulator is composed of two main components: a serving scheduler and a cluster.
To support continuous batching, we implemented a serving scheduler that manages ongoing inference requests.
The cluster receives device specifications and system configurations; then, it generates device components.
Based on the model distribution methodology, the simulator distributes model weights across these device components.
The operation of our simulator proceeds as follows:
1) The serving scheduler generates information about the requests being processed at each stage (\eg prefilling or decoding of each request and the current sequence length) and sends them to the cluster.
2) Upon receiving the requests, each device component within the cluster executes the assigned operations and results execution times.
For \lowp, we have modified the Ramulator, specifically the DRAM controllers and internal DRAM behavior models, to enable simultaneous data reading from all banks in the target \bankbundle.
We used the timing parameters of HBM3~\cite{jedec-hbm3} to simulate memory operations in both \duplex and \gpu.
For computing units, the timing data is calculated considering the number and the frequency of the computing units.
The cluster additionally computes the communication time for data movement between devices considering the latency and bandwidth of the HGX system~\cite{nvidia-hgx}, and based on the execution times received from the device components, calculates the final execution times.

We used \mixtral~\cite{arxiv-2024-mixtral}, \glam~\cite{icml-2022-GLaM}, \grok~\cite{grok1}, \opt~\cite{arxiv-2022-opt}, and \llama~\cite{2024-llama3} LLMs for evaluation. 
Mixtral and Grok1 have a structure with all MoE decoder blocks, while GLaM alternates decoder and MoE decoder blocks.
In the MoE layer, Mixtral and Grok1 select two out of eight experts, and GLaM selects two out of 64 experts per token.
To evaluate the performance of \duplex in conventional LLMs without MoE, we also used \opt and \llama.
\mixtral, \grok, and \llama uses GQA, and \glam and \opt uses MHA.
We used FP16~\cite{openai-fp16 ,arxiv-2022-opt} for weight precision.
Considering the number of model parameters, we configured the default number of nodes and the number of devices within each node as follows; \mixtral, \opt, \llama: one node with four devices, \glam: one node with eight devices, \grok: two nodes, each with eight devices.
Unless specified otherwise, we applied the data and job distribution method described in Section~\ref{sec:motivaiton}. 
For \doublegpu, we first increased the number of devices per node to a maximum of eight and increased the number of nodes.
Details of the model configurations are summarized in Table~\ref{tab:model_configuration}.

\begin{figure*}[!tb]
  \center
\includegraphics[width=1.00\textwidth]{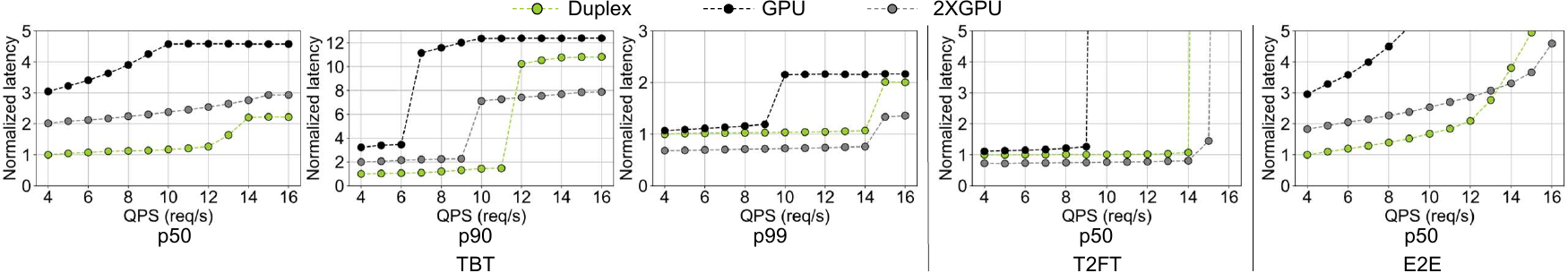}
  \vspace{-0.1in}
  \caption{
  The normalized latency (\tbt, \ttft, \ete) of \duplex, \gpu, and \doublegpu  for \mixtral varying queries per second (4 to 16). (\lin, \lout) is (4096, 512) and the maximum batch size is 128.
  }
  \label{fig:eval_qps}
  \vspace{-0.17in}
\end{figure*}

We used synthesized datasets to quantify the performance improvements of \duplex. 
We sampled the input and output lengths of each request using Gaussian distributions and represented the average of the input and output lengths for the sampled requests.
For expert selection, we chose the target experts for each token using a uniform distribution~\cite{jmlr-2022-switchtransformer}. 
To evaluate the performance in varying queries per second (QPS) situations, we injected requests into the systems following Poisson distributions~\cite{osdi-2022-orca, isca-2024-splitwise, ics-2024-clay} in the experiments shown in Fig.~\ref{fig:eval_qps}.
Otherwise, we assumed that when the inference of a request is finished, the next request is added to the batch and processed together in the next stage using continuous batching.

To measure area overheads and energy consumption, we synthesized the major components of \duplex devices.
We implemented arithmetic units in Verilog and synthesized them using Synopsys Design Compiler with a 7 nm predictive process design kit~\cite{asap7-2016}.
We set the operational frequency of arithmetic units 
of \highp as 1 GHz and \lowp as 650 MHz considering tCCDS of HBM3, which is 1.5 ns.
We modified FinCACTI~\cite{fincacti-2014} to match the published data of SRAM~\cite{isscc-2017-7nm-sram, whitepaper-2018-irds,vlsit-2018-samsung, isca-2021-tpuv4i, isscc-2018-7nm-sram-euv, iedm-2016-tsmc7nm} and used it to model the energy of SRAM-based buffers.
We referred to~\cite{MICRO-2017-fine-grained-dram} for the activation, read, write, and TSV energy of HBM.
We adjusted the area overhead for processing units and buffers on the DRAM die to 1z-nm DRAM technology~\cite{jssc-2022-hynix-hbm3, jssc-2023-samsung-hbm3}.
We then scaled the area overhead by factoring in that the DRAM process has 10$\times$ larger than the logic process for the same feature size~\cite{hotchips-2019-upmem}.

\section{Evaluation}
\label{sec:evaluation}

\subsection{Throughput Improvement of \duplex}
\label{sec:eval:duplex_throughput}
\duplex shows higher throughput than \gpu systems, with even \doublegpu in most cases by efficiently performing low~\opb MoE layers and attention layers using \lowp.
Fig.~\ref{fig:speedup_graph} shows the normalized throughput (tokens per second) of \duplex for various batch sizes and $(L_{in}, L_{out})$ configurations on three models compared to the \gpu.
To verify the performance enhancement of \duplex, we categorized \duplex into three configurations. 
\duplex is a device that uses only one of \highp or \lowp at any given time, as shown in Fig.~\ref{fig:duplex_operation_flow}(a) and (b). 
\duplexpe applies expert and attention co-processing, illustrated in Fig.~\ref{fig:duplex_operation_flow}(d). 
A device that incorporates tensor parallelism for MoE layers, as described in Section~\ref{sec:duplex_co-processing}, is referred to as \duplexpet.

\duplex already achieves up to 2.51$\times$ performance improvements compared to the baseline \gpu system, even showing the best performance for the $(L_{in}, L_{out}) =(4096, 4096)$ case of \mixtral.
There exist cases that \duplex outperforms the throughput of \doublegpu as \duplex utilizes greater memory bandwidth than that of \doublegpu in the low-\opb operations, which dominates the total execution time.

When only co-processing is applied, we observe an 1.04$\times$ on average in throughput compared to \duplex. 
Because each device processes fewer experts (two in the case of \mixtral and one for \grok), the benefits of expert co-processing are minimal.
While attention co-processing also reduces the latency of \mixedstage, it does not significantly improve throughput as \deconlystage dominates the stages in LLM inference.
\duplexpet enhances the effects of expert co-processing by employing tensor parallelism for experts as well, which increases the number of experts processed on each device, and increases throughput up to 1.36$\times$ and 2.67$\times$ compared to \duplex and \gpu.

\begin{figure}[!tb]
  \center
\includegraphics[width=1.0\columnwidth]{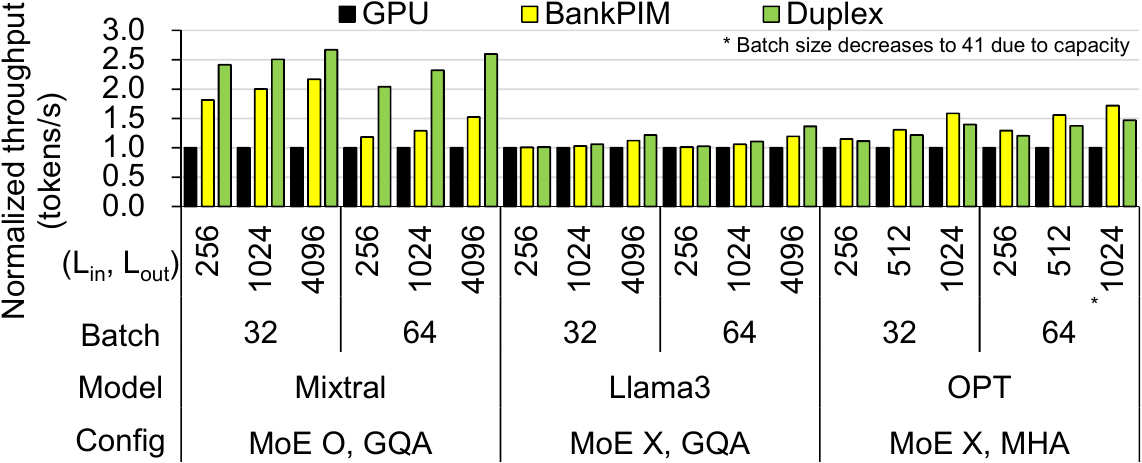}
  \vspace{-0.15in}
  \caption{
  The normalized throughput of \duplex and \bankpim in \mixtral, \llama, and \opt for various (\lin, \lout) from (256, 256) to (4096, 4096) and batch sizes (32--64).
  }
  \label{fig:eval_ablation_study_pim}
  \vspace{-0.1in}
\end{figure}

In large systems, performance improvements may be limited due to communication overhead between devices and nodes.
\grok exhibits smaller performance improvements compared to the other models. 
This is due to \grok's larger model size, which necessitates using two nodes for LLM inference. 
Relatively low bandwidth between nodes increases communication overhead, consequently diminishing the acceleration benefits for the MoE layer and attention layer using \duplex.

\begin{figure*}[!tb]
  \center
\includegraphics[width=1.0\textwidth]{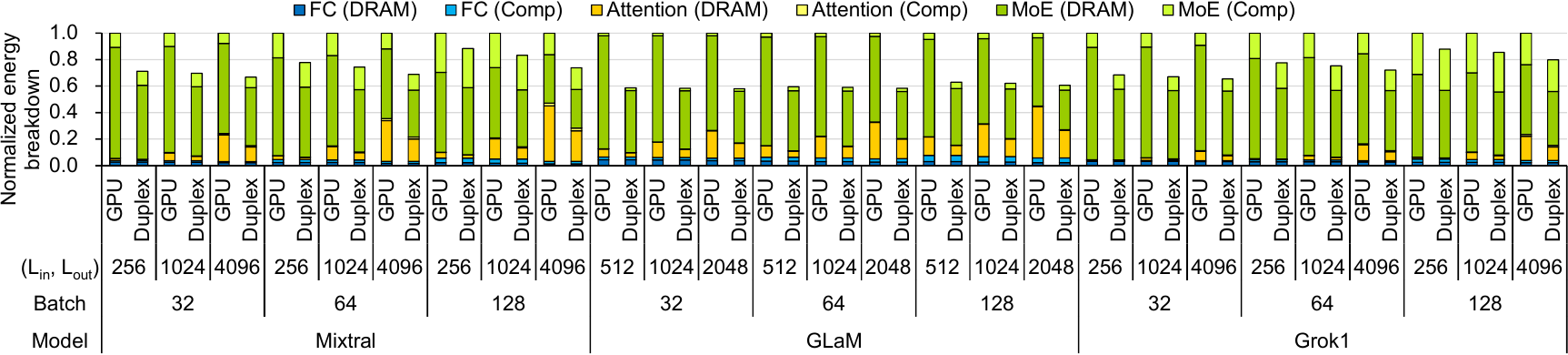}
  \vspace{-0.15in}
  \caption{
  The normalized energy breakdown of \mixtral, \glam, and \grok for various (\lin, \lout) from (256, 256) to (4096, 4096), and batch sizes (32--128).
  }
  \label{fig:energy_breakdown}
  \vspace{-0.12in}
\end{figure*}

\subsection{Latency Improvement of \duplex}
\label{sec:eval:duplex_latency}
\duplex significantly reduces the various types of latencies (\tbt, \ttft, \ete) over \gpu.
Fig.~\ref{fig:eval_latency_bar_graph} shows normalized latencies for \lin and \lout from 512 to 2048 in \glam with a batch size of 64.
On average, \duplex reduces the median \tbt value by 58.3\% by decreasing the execution times of the MoE layer and attention layer using \lowp compared to \gpu.
Further, \duplex achieves even lower median \tbt latency than \doublegpu.
The stage corresponding to the median \tbt latency is the \deconlystage, where the low~\opb MoE and attention layers dominate the execution time.
While \doublegpu utilizes twice as many processing units as \duplex to process the high-\opb FC layer quickly, \duplex can utilize twice the memory bandwidth as \doublegpu for low-\opb operations using \lowp.
By exploiting higher memory bandwidth using \lowp when processing dominant low~\opb operations in the \deconlystage, \duplex achieves a lower median \tbt latency than \doublegpu.

Even for median latencies of \ttft and 99th percentile for \tbt, which primarily occur in the \mixedstage well-suited to GPUs, \duplexpet achieves competitive latency improvements compared to \doublegpu.
When \lin is 512, \duplexpet could decrease the 99th percentile of \tbt and \ttft latencies up to 16.74\% and 26.17\% compared to \doublegpu. The \opb of the MoE layer in the \mixedstage is low enough to be accelerated by \lowp, making the expert and attention co-processing more effective.
When \lin is 2048, the \opb of the MoE layers in the mixed stage increases and the \lowp suffers from processing experts due to fewer processing units; thus, \duplexpet shows similar 99th percentile \tbt and \ttft with the \doublegpu.
By efficiently handling both \deconlystage and \mixedstage, \duplexpet reduces \ete latency by an average of 60.20\% and 35.38\% compared to \gpu and \doublegpu.

To evaluate the performance of \duplex under different serving intensities, we measured the latency of \duplex, \gpu, and \doublegpu with varying QPS (see Fig.~\ref{fig:eval_qps}).
\duplex always exhibits better median \tbt latency than \doublegpu.
Because the median \tbt latency is generally achieved during the \deconlystage, \duplex outperforms \doublegpu by exploiting higher memory bandwidth using \lowp compared to \doublegpu.
At low QPS, \duplex outperforms \doublegpu in the 90th percentile of \tbt latency.
However, as QPS increases, \duplex shows higher latency compared to \doublegpu.
For high QPS, as the system processes more \mixedstage{s}, \doublegpu performs the \mixedstage better by utilizing twice as many computing units as \duplex, lowering tail \tbt latencies over \duplex.
If requests exceed the system's throughput, \ttft latency skyrockets due to the queuing delay.
\gpu cannot handle requests if more than 9 requests are injected per second.
\duplex, which processes the \deconlystage faster, can handle up to 14 requests per second, nearly equivalent to the capability of \doublegpu.
Thus, \duplex always outperforms \gpu and demonstrates similar or better performance across various QPS.

\subsection{Comparison with \bankpim Across Various LLMs}
\label{sec:eval:compare_pim}
\duplex outperforms \duplexbankpim by efficiently accelerating low-\opb (over 1) operations (see Fig.~\ref{fig:eval_ablation_study_pim}) in LLM with MoE and GQA.
\bankpim shows up to 2.17$\times$ higher throughput than \gpu in the batch size of 32 when (\lin,\lout) is 4096.
When the batch size decreases from 64 to 32, the \opb of the MoE layer is lowered, leading to relatively respectable performance improvements in \bankpim.
As the batch size increases, the processing units in \bankpim struggle with processing the MoE layers due to increased \opb.
\bankpim cannot efficiently process MoE layers when the batch size increases, leading to diminished performance gains and showing only 1.18$\times$ higher throughput compared to \gpu when (\lin,\lout) is 256 with a batch size of 64.
\duplex exploits \lowp equipped with more processing units than \bankpim, exhibiting 2.05$\times$ higher throughput than \gpu in the same configuration.
As the \deggrp of \mixtral is 4, \bankpim shows similar speedups compared to \duplex in processing the attention layer of decoding sequences, despite having higher amplified bandwidth.
By providing more adequate memory bandwidth and processing units than \bankpim, \duplex shows up to 1.80$\times$ higher throughput and 1.49$\times$ higher throughput on average compared to \bankpim in \mixtral.

\duplex still achieves acceptable speedups compared to \gpu in out-of-target models, such as conventional LLM models without MoE layers.
As the attention operations of decoding sequences are low~\opb operations, \duplex shows performance improvement. 
In \llama, which uses GQA (\deggrp = 8), \duplex performs better than \bankpim.
While \duplex can efficiently handle GQA, \bankpim suffers from a lack of computing units.
In the case of \opt, which utilizes MHA, \bankpim performs better than \duplex.
Because the \opb of MHA in the \deconlystage is extremely low, \bankpim processes the attention layer of the \deconlystage faster than \lowp by utilizing high internal memory bandwidth, leading to higher throughput than \duplex.
\begin{figure}[!tb]
  \center
\includegraphics[width=1.0\columnwidth]{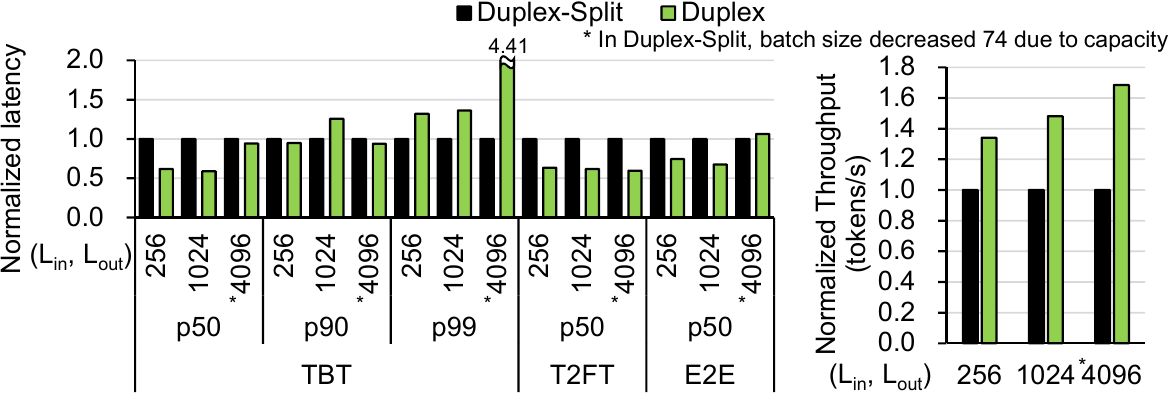}
  \caption{
  The normalized latency (\tbt, \ttft, \ete) of \mixtral for various (\lin, \lout) from (256, 256) to (4096, 4096) with a batch size of 128.
  \duplex and \duplexsplit each use a total of 4 \duplex devices. \duplexsplit processes each prefill stage and decoding stage with two \duplex units each.
  }
  \label{fig:split_duplex}
  \vspace{-0.15in}
\end{figure}

\subsection{Energy Consumption}
\label{sec:eval:energy}
\duplex reduces energy consumption by up to 33.28\%, 42.03\%, and 34.59\% for \mixtral, \glam, and \grok compared to \gpu.
Fig.~\ref{fig:energy_breakdown} shows normalized energy consumption for generating one token with \duplex and \gpu.
We can see that most of the energy is consumed in the MoE and attention layers.
\duplex reduces off-chip memory access energy in MoE and attention layer by leveraging \lowp.
The energy consumed by the attention layer rises as the sequence length increases, \duplex shows better energy efficiency in the long sequence lengths. 

In particular, energy efficiency deteriorates in \mixtral and \grok compared to \glam as the batch size increases. 
While \duplex reduces latency through co-processing, it consumes more energy by utilizing \highp, which uses more DRAM energy than \lowp. 
As \mixtral and \grok employ fewer experts than \glam, resulting in a higher \opb for the MoE layers in \mixtral and \grok than in \glam at the same batch size.
Thus, \duplex relies on \highp to process more experts in \mixtral and \grok compared to \glam, leading to relatively lower energy efficiency compared to \gpu when the batch size increases.

\subsection{Area Overhead}
\label{sec:eval:area_overhead}
The total overhead for processing units in \duplex for each \lowp stack is 17.80 mm$^2$, which accounts for 14.71\% of a 121 mm$^2$ HBM3 logic die~\cite{jssc-2022-hynix-hbm3}. 
The pitch of TSV was measured at 22 um~\cite{SKhynix-tsv-pitch}, and the number of TSVs per channel was conservatively scaled by increasing it to four times the number required per channel in conventional HBM3~\cite{SKhynix-tsv-pitch}. 
The added TSVs account for an area overhead of 10.89 mm$^2$. 
Each \lowp stack includes 32 GEMM modules, with each module comprising 512 FP16 MACs operating at 650MHz and a 8KB buffer, accounting for 3.02 mm$^2$.
Further, \lowp contains two 1MB buffers to store input vectors and temporal results, occupying 2.26 mm$^2$.
A softmax unit, which consists of a comparator tree, adders, exponential units, an adder tree, dividers, multipliers, and a total of 128 KB buffers, occupies 1.64 mm$^2$.
Considering that the area overhead ranges from 20\% to 27\%~\cite{isca-2021-hbm-pim, isscc-2022-aim}, \duplex demonstrates significant performance improvements with a lower area overhead.

\section{Discussion}
\label{sec:discussion}

\subsection{Split Prefill and Decoding Node}
\label{sec:discussion:splitwise}

Splitwise~\cite{isca-2024-splitwise} proposed an LLM inference system by dividing nodes into prefill (prompt) and decoding (token) nodes. 
Based on the importance of high computing power for the prefill stage and memory bandwidth for the decoding stage, Splitwise suggested a cost-effective system by deploying suitable hardware across prefill and decoding nodes. 
This approach benefits in low tail latency of \tbt compared to non-split systems, as no mixed stages are involved in token generation within the decoding nodes.

However, split systems show lower throughput than non-split systems due to the underutilization problems and the wasted memory capacity due to weight duplication (see Fig.~\ref{fig:split_duplex}).
The utilization of prefill and decoding nodes varies across batch sizes, input, and output sequence lengths. 
One of the prefill or decoding nodes would suffer from low utilization unless it targets a specific scenario. 
For cloud service providers, managing separate devices for prefill and decoding stages and reconfiguring the system to prevent underutilization for each target scenario is highly burdensome. 
Further, the split approach incurs memory weight duplication, limiting batch size due to wasted memory capacity and thus degrading throughput compared to a non-split system.

\subsection{Implications of Expert Skews on Expert Co-processing}
\label{sec:discussion:expert_distribution}
A prior work~\cite{2023-arxiv-moe-skew} argue that the number of tokens each expert processes can substantially differ. 
In cases where there are hot experts processing a large number of tokens and cold experts processing a smaller number of tokens, \duplex can efficiently process the MoE layer by flexibly handling experts with both \highp and \lowp, exploiting expert co-processing. 
However, in ideal cases where each expert processes the same number of tokens, expert co-processing may not be as effective.

\subsection{KV Cache Migration and Recomputation}
\label{sec:discussion:kvcache}
The generated KV cache size increases in proportion to the batch size and sequence length. 
With the current trend of increasing sequence length~\cite{cal-2023-attacc}, the KV cache could lead to a shortage of device memory capacity in a system. 
PagedAttention~\cite{sosp-2023-pagedattention} proposed KV cache migration and recomputation when its capacity exceeds the memory capacity of the GPU system.
KV cache migration involves evicting a portion of requests and migrating the KV caches to CPU memory to free up device memory capacity.
Once the inference of the ongoing requests are completed and there is available device memory capacity, the system brings the evicted KV caches back to the device memory to resume the inference. 
In the recomputation method, the KV caches are deleted instead of being migrated.
When the inference of evicted requests is resumed, the previous KV caches are recomputed.
These methods can be complementarily applied to \duplex.

\section{Related Work}
\label{sec:related_work}
Most ASIC- and FPGA-based LLM accelerators have focused on quantization or pruning to reduce the size of the model or memory usage.
\cite{hpca-2020-a3, isca-2021-elsa, hpca-2021-spatten, micro-2021-sanger, isca-2023-fact, asplos-2022-dota, isca-2022-mokey, isca-2022-leopard, micro-2022-butterfly, vlsi-2022-nmsparse} focused on the presence of unnecessary values in the attention matrix, which is the result of multiplying the Q and K matrices within the MHA layer of the transformer model. 
They proposed accelerators that approximate attention operations or prune unnecessary values in the attention matrix through methods such as eager prediction.
\cite{micro-2020-gobo, hpca-2021-spatten, isca-2023-olive} proposed accelerators that support quantization of the weight or embedding vector of the transformer model.
Meanwhile, \cite{micro-2022-dfx, asplos-2023-flat} proposed accelerators that process the transformer model losslessly.
\cite{micro-2022-dfx} proposed a system that distributes processing across multiple FPGAs for low-latency inference of the transformer model and \cite{asplos-2023-flat} proposed a dataflow for the accelerator to improve the efficiency of the attention operation.

Numerous studies have proposed leveraging near-memory processing or PIM to process the transformer model effectively.
\cite{isca-2021-hbm-pim, isscc-2022-aim} introduced commercialized PIM architectures on HBM and GDDR, respectively.
They proposed a system and software stack that offloads the GEMV operations of the transformer model to the PIM architecture with in-bank processing units.
\cite{asplos-2024-attacc} proposed a PIM architecture that efficiently handles attention operations with extremely low-\opb and high memory capacity requirements on large batch sizes.
\cite{asplos-2024-neupim} proposed the integration of NPU and PIMs to efficiently handle both GEMM and GEMV.
\cite{hpca-2022-transpim} proposed an end-to-end accelerator to process the transformer model on PIM.

TOP-PIM~\cite{hpdc-2014-toppim} proposed GPU with PIM stacks, achieving acceleration for memory-intensive workloads through the internal stack bandwidth between the DRAM die and logic die, which is higher than the I/O bandwidth of the HBM stack.
However, it did not aim to accelerate LLMs and lacks details of DRAM die architectures.
Also, it did not consider concurrently executing GPUs and PIM stacks; thus, the expert and attention co-processing cannot be applied.
Despite these works on accelerating LLMs and PIM architectures, to the best of our knowledge, \duplex is the first to accelerate LLMs with MoE and GQA in continuous batching.

\label{sec:related_work:accelerator}

\section{Conclusion}
\label{sec:conclusion}
In this paper, we have proposed \duplex, a device to efficiently accelerate large language models (LLMs).
We observed that low-\opb operations dominate the total execution time in LLMs and that utilizing heterogeneous systems following prior work would result in severe performance degradation.
\duplex integrates two types of processing units that share device memories, each designed to efficiently handle high-\opb and low-\opb layers in an LLM.
Especially, we introduced an alternative processing-in-memory (PIM) microarchitecture, \lowp, that exploits the increased TSV density of contemporary HBM to place powerful processing units on the logic die of an HBM.
We proposed expert and attention parallel execution to maximize the utilization of processing units in \duplex.
By these means, \duplex shows up to 2.67$\times$ higher throughput and 42.03\% less energy consumption, achieving 2.07$\times$ higher throughput and 28.19\% lower energy consumption on average for LLM inference compared to a GPU system using H100.

\section*{Acknowledgment}
This work was partly supported by the Samsung Electronics, the National Research Foundation of Korea grant funded by the Korea government (MSIT) (RS-2024-00405857), Institute of Information \& communications Technology Planning \& Evaluation (IITP) grant funded by MSIT (RS-2024-00402898, RS-2021-II211343), and IITP under the artificial intelligence semiconductor support program funded by MSIT (IITP-2023-RS-2023-00256081).
The EDA tool was supported by the IC Design Education Center (IDEC), Korea.
The ICT at Seoul National University provides research facilities for this study.
Jung Ho Ahn is the corresponding author.


\balance
\bibliographystyle{IEEEtranS}
\bibliography{refs}

\end{document}